\documentclass[aps,preprint,amssymb]{revtex4}
\usepackage{epsfig,color}
\def\MoSb{Mo$_3$Sb$_7$}

\begin{document}

\title{
Low-temperature specific heat of the superconductor \MoSb
}
\author{V. H. Tran$^1$, W. Miiller$^1$, Z. Bukowski$^2$}
\affiliation{$^1$ W. Trzebiatowski Institute of Low Temperature and
Structure Research, Polish Academy of Sciences, P.O. Box 1410, 50-950 Wroc\l
aw, Poland\\
$^2$ Laboratory for Solid State Physics, ETH Z\"{u}rich, 8093 Z\"{u}rich, Switzerland}

\begin{abstract}
The low-temperature specific heat of a superconductor \MoSb~with $T_c$ = 2.25 $\pm 0.05$ K has been measured in magnetic fields up to 5 T. In the normal state, the electronic specific heat coefficient $\gamma_n$, and the Debye temperature $\Theta_D$ are found to be 34.5(2) mJ/molK$^2$ and 283(5) K, respectively. The enhanced $\gamma_n$ value is interpreted due to a narrow Mo-4d band pinned at the Fermi level. The electronic specific heat in the superconducting state can be analyzed in terms a phenomenological two BCS-like gap model with the gap widths 2$\Delta_1/k_BT_c$ = 4.0 and 2$\Delta_2/k_BT_c$ = 2.5, and relative weights of the mole electronic heat coefficients $\gamma_1/\gamma_n$ = 0.7 and $\gamma_2/\gamma_n$ = 0.3. Some characteristic thermodynamic parameters for the studied superconductor, like the specific heat jump at $T_c$, $\Delta C(T_c)/\gamma_n T_c$, the electron-phonon coupling constant, $\lambda_{e-ph}$, the upper $H_{c2}$ and thermodynamic critical $H_{c0}$ fields, the penetration depth $\lambda$, coherence length $\xi$  and the  Ginzburg-Landau parameter $\kappa$ are evaluated. The estimated values of parameters like 2$\Delta_0/k_BT_c$, $\Delta C(T_c)/\gamma_n T_c$, $N(E_F)$, and $\lambda_{e-ph}$ suggest that \MoSb~belongs to intermediate-coupling regime. The electronic band structure calculations indicate that the density of states near the Fermi level is formed mainly by the Mo-4d orbitals and there is no overlapping between the Mo-4d and Sb-sp orbitals.

\end{abstract}
%


\maketitle

\hyphenation{tem-per-at-ure}

\section{Introduction}
Several years ago, Bukowski and coworkers reported the superconductivity in an intermetallic \MoSb \cite{bukowski}.
The superconducting phase transition of this material has been determined from the magnetization and electrical resistivity on a
single crystal to be 2.08 K \cite{bukowski}. Based on the magnetization \cite{bukowski},  and point-contact (PC) Andreev-reflection data \cite{dmitriev1}, the authors of Ref. \cite{bukowski,dmitriev1} concluded the BCS-type behaviour of the superconductor.  However, in the recent work,  Dmitriev et al.\cite{dmitriev}  had suggested that \MoSb~is not a trivial BCS (s-wave) superconductor but rather it has a (s+g)-wave or another unconventional pairing symmetry. More recently, Candolfi et al. \cite{candolfi}, have reported magnetic susceptibility, specific heat, and electrical resistivity down to 0.6 K and have interpreted their data within the frameworks of the spin fluctuation theory. Thus, the issue of pairing mechanism becomes debated and the proper superconductivity nature of \MoSb~deserves studies in more details. Nonetheless, information about the superconducting
 properties of \MoSb~is still scarce. Except for the superconducting transition, other parameters like upper $H_{c2}$ and thermodynamic critical $H_{c0}$ fields, the penetration depth $\lambda_{L}$, coherence length $\xi$  and the  Ginzburg-Landau parameter $\kappa$ are not known yet. In order to
 determine these fundamental thermodynamic parameters of the \MoSb~superconductor, we have performed specific heat $C_p$ at temperatures 0.4 -10 K and in magnetic fields up to 5 T. In this work, we report the results of the $C_p$ measurements and we try also to address the question: which electron pairing model is more suitable for description of the superconductivity in \MoSb? The specific heat data are supplemented by electronic band structure calculations performed using the the full-potential local-orbit method.

 \section{Experiments}
  \par A powdered sample of \MoSb~was prepared from mixture of powder metals Mo and Sb with
  purities of at least 99.95\% from Alfa Aesar by annealing in an evacuated silica tube at 700$^\circ$ for three days.
  The mixture was then grounded and pressed into the form of cylinder, and annealed at 600$^\circ$ for two days.
  The purity, homogeneity and composition of the sample were checked by X-ray diffraction and dispersive spectroscopy.
  \par Fig. \ref{Xray} shows the experimental powder X-ray diffraction patterns of \MoSb~and as well as the profile fit and difference patterns in the Rietveld analysis. It can be seen, the observed Bragg reflections were indexed on the base of the cubic Ir$_3$Ge$_7$-type structure (space group Im-3m) \cite{Im3m}. In the refinements, we assumed that Mo atoms occupy the 12e (x = 0.342) and the antimony distribute on the 12d and 16f (x = 0.162) sites, respectively. The refinements of the X-ray intensities yield lattice parameter \emph{a} = 0.9551 (1) nm, in well agreement with that previously reported by Dashjav et al. for a single crystal (0.9559 nm) \cite{Dashjav}.
\par The microanalysis was performed by collecting several EDX spectra at various locations across the surface of a sintered pellet. The compositions determined by EDX Mo:  29.2 $\pm~3~at. \%$ and Sb: 70.8 $\pm~3~at. \%$ are close to the ideal composition Mo:Sb= 3:7. No impurity phase within the experimental error is observed in the EDX spectra. The SEM image of the typical samples is shown in the inset of Fig. \ref{Xray}.
   \par The specific heat was measured using a thermal relaxation method in the temperature range 0.4 - 10 K
  and in magnetic fields up to 5 T, utilizing PPMS system (Quantum Design). The experimental errors were taken as found by Marriott et al. \cite{PPMS}, i.e., to be less than $\pm$ 2 \% for T$>$ 1.5 K and about $\pm$ 20\% at 0.4 K. Before the specific heat study, the bulk superconductivity in the investigated sample was examined by means of magnetization and electrical resistivity measurements. From the latter measurement we obtained the residual resistivity in the normal state $\rho_n$ = 95x10$^{-8}$ $\Omega$m.
  \par The electronic band structure calculations were carried out using
the full-potential local-orbit method (FPLO-5) \cite{Kopernik1,Kopernik2,web}. The
calculations were performed for the reciprocal space mesh containing 816 points within the
irreducible wedge of the Brillouin zone.

\section{Results}

\par The calculated densities of states (DOS) of \MoSb~near the Fermi level are shown
in Fig. \ref{DOS}, together with the partial contributions from the constituent Mo and Sb atoms. The global feature of the DOS is in reasonable agreement with the results of Self-consistent LMTO (linear muffin tin orbitals) calculations performed by Dashjav et al \cite{Dashjav}. Both calculations have indicated the dominating contribution of the Mo-4d orbitals to the total DOS. The calculated DOS at the Fermi level $N(E_F)$ is estimated to be 42 states/eV unit cell, corresponding to about 1.04 states/eV atom. Comparison of the obtained $N(E_F)$ value with those of YBa$_2$Cu$_3$O$_7$ and MgB$_2$ ($\sim$ 0.12 states/eVatom.spin  \cite{Junod}-\cite{Wang}) points to an enhancement of the $N(E_F)$ of \MoSb. This observation implies that the 4d-electrons of the Mo atoms forming a narrow band  at the Fermi level give rise to the density of states, in similar manner as in the case of the A15 superconductors \cite{A15}. A distinct difference between our and the previous published data lies in the fact that the position of the Fermi level in our calculations is shifted toward higher energy by 0.02 eV. Thus, $E_F$ situates in the opposite site of a local valley. The important result emerging from our calculations is not only the fact that the largest peak of DOS located at -0.035 eV below $E_F$ (band width $E_B$ of $\sim$ 0.05 eV) forms mainly from the Mo-4d orbitals but also no overlapping between Mo-4d and Sb-sp orbitals is observed. Thus, the hybridization of the Mo-4d with the valence sp-band of the Sb atoms seems to be negligible and the Mo-4d and Sb-sp orbitals contribute to the total DOS probably separately.
We recognize also that the contribution of the sp-band of the Sb2 atoms to the total DOS at $E_F$ is much lower than that of Sb1. This finding may support a very weak hybridization strength between Mo-4d and Sb-sp orbitals, since the distance Mo-Sb2 ($\sim$ 2.78 \AA) is shorter than the Mo-Sb1 ($\sim$ 2.82 \AA) one.

 \par The specific heat $C_p$ data of \MoSb~at zero field is shown in Fig. \ref{CpT} in the form $C_p/T$ vs. $T^2$. The occurrence of superconductivity is evidenced by a specific heat jump at 2.2 $\pm$0.05 K. The superconducting phase transition temperature determined from the specific heat measurement is consistent with those obtained from the magnetization and electrical resistivity (not shown here) and with the specific data reported by Candolfi et al. \cite{candolfi}. Note,  the value of $T_c$ observed in the present sample is of about 0.1 K higher than that found in a single crystal of 2.08 K, \cite{bukowski}, though  specific heat data for single crystal are not yet available. The difference in the superconducting transition values between the single-crystalline and polycrystalline samples seems to be associated with the change in the densities of states at the Fermi level. Since the unit cell volume of the studied sample with 871.34 \AA$^3$ is smaller than that of the single-crystalline sample (879.769 \AA$^3$) \cite{bukowski}, one supposes that the decrease in the volume results in displacement of the 4d-level to a higher energy respect to $E_F$, and in consequence increases DOS.
\par The normal-state specific heat data of \MoSb~can be analyzed using the standard formula:
\begin{equation}
C_p/T = \gamma_n + \beta T^2,
\label{eq.1}
\end{equation}
 where the first term gives the electronic contribution and the second represents the lattice contribution to the total specific heat. For the temperature range 2.5 - 10 K, it is found that $\gamma_n$ = 34.5(2) mJ/molK$^2$ and $\beta$ = 0.85(5) mJ/molK$^4$. The latter value corresponds to the Debye temperature of 283(5) K.

\par From the specific heat data (Fig. \ref{CpT}), the dimensionless specific heat jump $\Delta C(T)/\gamma_n T_c$ was evaluated to be 1.56. Thus, there is a substantial difference in the $\Delta C(T)/\gamma_n T_c$ values obtained by us and by Candolfi et al. \cite{candolfi} i.e., 1.04.  Since, the value of 1.43 is the weak-coupling limit \cite{BCS}, the superconductivity in the sample studied here could belong to the strong-coupling regime. An information about the strength of the electron-phonon coupling can be gained by evaluating the average electron-phonon coupling constant $\bar{\lambda}_{e-ph}$ from the McMillan equation \cite{McMillan}:
 \begin{equation}
\lambda = \frac{1.04+\mu^*ln(\Theta_D/1.45T_c)}{(1-0.62\mu^*)ln(\Theta_D/1.45T_c)-1.04}
\label{McMillan}
\end{equation}
where $\mu^*$ is the Coulomb pseudopotential and it is usually taken between 0.1 and 0.15 \cite{McMillan,MA}.
Putting $\mu^*$ = 0.15, $\Theta_D$ = 283 K and $T_c$ = 2.2 K into Eq. \ref{McMillan} we obtained $\lambda_{e-ph}$ = 0.57. For \MoSb~the value of $\mu^* \approx$ 0.17 can be roughly estimated from the electronic band gap $E_B$ = 0.05 eV, phonon frequency $\omega_0 \approx \Theta_D$ = 283 K and product $N(E_F)V \approx$ 0.2 using the equation:
$\mu^* = \frac{N(E_F)V}{1+N(E_F)Vln(E_B/\omega_0)}$ \cite{MA}. The value of the product $N(E_F)V$ was calculated with help of the BCS equation: $T_c = 1.14\Theta_D exp(-1/N(E_F)V)$.

\par The knowledge of $\gamma_n$ allows us to estimate the electron-phonon mass enhancement $\lambda_{\gamma}$ parameter,  which should be similar in magnitude to $\lambda_{e-ph}$. Employing the relation $\gamma = \frac{2}{3}\pi^2N_Ak_B^2(1+\lambda_{\gamma})N(E_F)$ and taking $N(E_F)$ as the DOS obtained by the electronic band structure calculations, we obtained $\lambda_\gamma$ = 0.64. Compared to weak-coupling $\lambda_{e-ph}$ value of 0.4, the relatively large and consistent values of $\lambda_{e-ph}$ and $\lambda_{\gamma}$ suggest an intermediate strong coupling strength. This finding is consistent with an enhanced value of the specific heat jump at $T_c$.
\par We derived the electronic specific heat $C_{el} =  C_p - \beta T^3$ and have plotted the data in Fig. \ref{Cel} a in the form of semi-logarithmic scale $ln(C_{el})$ vs 1/\emph{T}. Clearly, in the superconducting state, the $C_{el}(T)$ curve reflects an exponential behaviour, indicating a gap character of the studied superconductor.  We wish to add that within the experimental error we have not seen any power law $T^2$ or $T^3$ in the temperature dependence of the electronic specific heat. This observation rules out any anisotropic symmetry of superconductivity as suggested by Dmitriev et al. \cite{dmitriev}. However, an attempt to fit the experimental data with the phenomenological relation for the conventional s-wave superconductors: $C_{el} = 8.5\gamma_n T_c exp(-1.44 T_c/T)$ \cite{gopal}, for the whole temperature range studied (represented as dashed line) is failed. In fact, the single-gapped BCS-like $C_{el} = A\gamma_n T_c exp(-\Delta_0/k_BT)$ function describes well the experimental data only in a limited temperature range, e.g, for 1 K $ <T < $ 2.2 K we obtained parameters \emph{A} = 12.7 and $\Delta_0/k_B$ = 4.0, corresponding to $2\Delta_0/k_BT_c \approx$ 3.63.
\par In order to reproduce the data in the whole temperature range studied, one applies a phenomenological two-gap model, which was proposed  by Wang et al. \cite{WPJ}, and by Bouquet et al. \cite{Bouquet} some years ago for MgB$_2$.
Following the latter authors, the electronic specific heat below $T_c$ can be considered as the sum of the contributions of two bands characterized by respective gap widths $\Delta_1$ and $\Delta_2$, and by partial Sommerfeld ratio $\gamma_1$ and $\gamma_2$, where $\gamma_n = \gamma_1 + \gamma_2$. We have fitted the experimental data to theoretical curves within the $\alpha$-model \cite{Padamsee}.
The theoretical data were calculated based on the following equations:
\begin{equation}
C_{el}(t)/\gamma T_c=t(d/dt)(S_{el}(t)/\gamma T_c)
\end{equation}
\begin{equation}
S_{el}(t)/\gamma T_c=-(3\alpha/\pi^2)\int_0^\infty [f(x){\rm ln}(f(x))+(1-f(x)){\rm ln}(1-f(x))]dx
\end{equation}
where   $\alpha$ is the gap ratio  $\alpha$ = $\Delta(T)/\Delta_{BCS}(T)$ with 2$\Delta_{BCS}(0)=3.5k_BT_c$ and  $f(x)= [{\rm exp}(\frac{\alpha (x^2+\delta(t)^2)^{1/2}}{t})+1]^{-1}$. $\delta(t)$ is the normalized BCS gap at the reduced temperature  $t = T/T_c$ and tabulated by M\"{u}hlschlegel \cite{Muhlschlegel}.
The best fitting result was observed for the following parameters 2$\Delta_1$ = 4.0$k_BT_c$ and  2$\Delta_2$ =2.5$k_BT_c$ and relative mole weight $\gamma_1/\gamma_2$ = 70/30.
The good fitting of the specific heat data to the exponential law implies that \MoSb~may be an s-wave-like superconductor with two energy gaps dependent on temperature.
The existence of two bands having different electronic densities at the Fermi level can be
understood if we assume that a narrow Mo-4d band and a broader Sb-sp band contribute
differently to $N(E_F)$. Based on the partial gap widths, one suspects that the larger gap
determined by the stronger electron-phonon coupling would correspond to the Mo-4d band,
while the smaller gap associates with the Sb-sp band, and then with $\gamma_n$ = 34.5 mJ/molK$^2$
and the relative $\gamma_1/\gamma_2$ ratio, one estimates the partial electronic coefficient $\gamma_2$ =1.5 mJ/(g
at.Sb K$^2$) and $\gamma_1$ = 8.0 mJ/(g at.Mo K$^2$). This viewpoint is quite accordable with the results
of the electronic band structure calculations shown above, where the dominating input of
the Mo atoms to the DOS is observed. We may add that two-band gap feature has been found in some superconductors, except for MgB$_2$ mentioned above, also in YbNi$_2$B$_2$C \cite{Huang} and Nb$_3$Sn \cite{Guritanu}.

 \par  The electronic entropy in the superconducting state was calculated through $S_{SC} = \int_0^{T_c} C_{el}/TdT$ (Fig. \ref{Cel} b).  At $T_c$,  $S_{SC}$ amounts to 80 mJ/molK and it is comparable with the entropy in the normal state $S_n = \gamma_n T_c$ of 76 mJ/molK assuming a constant $\gamma_n$ = 34.5 mJ/molK$^2$ below $T_c$ = 2.2 K.
\par The temperature dependence of the thermodynamic critical field $\mu_0H_{c0}$ is obtained by an integration of the data in the superconducting state according to the equation:
\begin{equation}
\frac{1}{2}\mu_0V_MH_{c0}^2(T)=\Delta U(T) -T\Delta S(T)
\label{dF}
\end{equation}
where $V_M$ is the volume per mole,  $\Delta U$ is the internal energy difference  and $\Delta S$ is the entropy difference.
From the equations:
\begin{equation}
\Delta U(T) = \int_T^{T_c}(C_{el}(T^\prime)-\gamma_nT^\prime)dT^\prime
\end{equation}

\begin{equation}
\Delta S(T) = \int_T^{T_c}\frac{C_{el}(T^\prime)-\gamma_nT^\prime}{T^\prime}dT^\prime
\end{equation}
 we computed the temperature dependence of $\Delta U$ and $T\Delta S$ (see inset of Fig. \ref{Hc0}).
The critical field $\mu_0H_{c0}$ is calculated from Eq. \ref{dF} is plotted in Fig. \ref{Hc0} as a function of temperature. An extrapolation to 0 K yields $\mu_0H_{c0}(0)$ = 0.029 T. A theoretical $\mu_0H_{c0}(0)$-value can be approximated from the relation: $\mu_0H_{c0}(0)$ = 7.65x10$^{-4}\gamma^{1/2}T_c$ \cite{Hake}, to be $\approx$ 0.027 T. This value is in good agreement with the experimental one. Using the relationship between the $\mu_0H_{c0}(0)$ and the superconducting energy gap $\Delta(0)$: $\mu_0V_MH_{c0}^2(0)= (\frac{3\gamma}{2\pi^2 k_B^2})\Delta(0)^2$ we deduce $\Delta(0)$ to be 0.39 meV ($\sim$ 4.18 K). From this value, the gap 2$\Delta(0)/k_BT_c$ for \MoSb~is calculated to be 3.8, slightly larger than  BCS value of 3.52 for weak-coupling superconductors.

\par In order to analyze the upper critical field $H_{c2}$, we measured specific heat at several selected magnetic field strengths (Fig. \ref{Cpfield}). With increasing magnetic fields up to 2 T, the superconducting transition shifts down to lower temperatures. The resulting upper critical field against temperature is shown in Fig. \ref{Hc2} a.
Apparently, the values of $\mu_0H_{c2}(T)$ can be described by a relation $\mu_0H_{c2}(T)$ = $\mu_0H_{c2}(0)[1 - (T/T_c)^{1.29}]$ with $\mu_0H_{c2}(0)$ = 2.24 T and $T_c$ = 2.2 K. Near $T_c$ the initial gradient $-\mu_0dH_{c2}(T)/dT$ amounts to 1.25 T/K. An independent determination of $-\mu_0dH_{c2}(T)/dT$ may be made using the equation given by Orlando et al. \cite{Orlando}: $-\mu_0dH_{c2}(T)/dT$ = 4.48x10$^3 \gamma \rho_0$ in the dirty limit. For the experimental values of $\gamma$ and $\rho_0$ we estimated $-\mu_0dH_{c2}(T)/dT \approx$ 1.12 T. Thus, there is satisfactory agreement with the experimental value.\\
\par Using the relationship $\mu_0H_{c2}(0) \approx 0.69 (-\mu_0dH_{c2}(T)/dT)T_c$ for a type-II superconductor in a dirty limit \cite{Werthamer}, we estimated $\mu_0H_{c2}(0)$ to be 1.9 T which is, however, slightly smaller than the value $\mu_0H_{c2}(0)$ fitted above. A better agreement between the theoretical and experimental values might have been obtained if one takes an additional parameter characterizing spin-orbital coupling $\lambda_{SO}$.
\par The value $\mu_0H_{c2}(0)$ can be estimated with the help of the Maki theory \cite{maki}. The upper critical field is given through the relation:  $\mu_0H_{c2}(0) = \alpha_M H_{po}/\sqrt{2}$, where the BCS value of the Pauli limiting field is $H_{po}^{BCS} = 1.84T_c$ and  $\alpha_M$ is the Maki parameter. Taking into account the electron-phonon corrections to the Pauli limiting process the field $H_{po}$ can be written as $H_{po} = H_{po}^{BCS} \sqrt{1+\lambda_{e-ph}}$ \cite{Clogston,Orlando}. The parameter $\alpha_M$ is obtained through the relation $\alpha_M$ = -0.528$\mu_0dH_{c2}/dT$ or alternatively $\alpha_M = 3e^2\hbar\gamma_n\rho_n/(2m\pi^2k_B^2)$, where \emph{m} is the mass of the free electrons. With $\mu_0dH_{c2}/dT$=-1.25, and $\gamma_n$ = 263 J/m$^3$K$^2$ and $\rho_n$ = 95x10$^{-8}$ $\Omega$m, the value of the Maki parameter amounts to 0.66 and 0.59, respectively. Using the average value $\alpha_M$ = 0.625 and $\lambda_{e-ph}$ =0.6 we obtain the upper critical field $\mu_0H_{c2}(0)$ = 2.26 T, which is close to the experimental value of 2.24 T.
\par Having upper and thermodynamic critical fields, we can estimate the type-II characteristic Ginzburg-Landau (GL) parameters at 0 K. From the relation:
$
\mu_0H_{c2}(0)=\frac{\Phi_0}{2\pi \xi_{GL}^2},
$
where $\Phi_0$ = 2.067x10$^{-15}$ Wb is the flux quantum we obtained coherence length $\xi_{GL} \approx$ 12 nm.
Using the relation: $\mu_0H_{c2}(0)=\sqrt{2}\kappa \mu_0H_c(0)$ we get Ginzburg-Landau parameter $\kappa$ = 54.6. The relation: $\mu_0H_{c}(0)=\frac{\Phi_0 \kappa}{2\pi \sqrt{2}\lambda^2}$ results in the penetration depth $\lambda \approx$  660 nm. These parameters can be evaluated by other different method with the help of relations given in Ref. \cite{Orlando}. From $\lambda$ = 6.42x10$^{-4} \rho /T_c$ and $\kappa$ = 2.37x10$^6 \gamma^{1/2} \rho$, the parameters $\kappa$ and $\lambda$ were calculated to be 420 nm and 36.5, respectively. We suppose that smaller values of these parameters deduced from the resistivity may be due to an underestimated value of $\rho$.

\begin{table}
\caption{Characteristic parameters of the superconductor \MoSb.}
\begin{tabular}{|l|c|}
  \hline
  $T_c$ (K) & 2.2(0.05)\\
  $\gamma_n$ [mJ/(molK$^2$)]& 34.5(2) \\
  $\beta$ [mJ/(molK$^4$)]&  0.85(5)\\
  $\Theta_D$ (K)& 283(5) \\
  $\lambda_{e-ph}$ & 0.6(0.05) \\
  $N(E_F)$ [states/(eV at.)]& 1.0 \\
  $\Delta C(T_c)/(\gamma_n T_c)$ & 1.56 \\
  2$\Delta_1/(k_BT_c)$ & 4.0 \\
  2$\Delta_2/(k_BT_c)$ & 2.5 \\
  -$\mu_0(dH_{c2}/dT)$ (T/K)& 1.25 \\
  $\mu_0H_{c2}$ (T) & 2.24 \\
  $\mu_0H_{c0}$ (T) & 0.029 \\
  $\xi_{GL}$ (nm)& 12 \\
  $\kappa$ & 36 - 55 \\
  $\lambda_{L}$ (nm)& 420 - 660\\
  \hline

\end{tabular}
\end{table}
\par The field dependence of the Sommerfeld coefficient at 0.4 K, $\gamma_{0.4 K}(H) \equiv C_p/0.4(H) $, is displayed Fig. \ref{Hc2} b. We observe linear dependence of the $\gamma_{0.4 K}$ with $\mu_0H$ up to 1.5 T, shown by the dashed line. The linearity of $\gamma_{0.4 K}(H)$ curve of \MoSb~accords with the theory for gapped superconductors \cite{CGM}. Although, such a linear behaviour of $\gamma_{0.4 K}(H)$ is not observed in fields up to $\mu_0H_{c2}$. Above 2 T the $\gamma_{0.4 K}(H)$ levels off to a constant. Thus, the spin-fluctuation scenario in \MoSb~is not confirmed by our specific heat data, where two characteristic features of spin fluctuations, namely $C_p(T) \sim T^3ln(T)$ \cite{SpinFluct}, and $\gamma(H) \sim H^2$ \cite{BMF}, are not found.
\section{Conclusions}
\par In the present work we have presented a detailed analysis of the low-temperature specific heat data of the superconductor \MoSb. The lattice specific heat is characterized by the Debye temperature of 283 K. The observed Sommerfeld coefficient of $\gamma_n$ = 34.5 mJ/molK$^2$ and a large electronic density of states, $N(E_F)$ of 1.04 states/eV atom, are indicative of a narrow band of the Mo atoms at the Fermi level. We have observed that the electronic specific heat in the superconducting state follows a single gapped function in a narrow temperature range. The gap width of 2$\Delta_0/k_BT_c \approx$ 3.64 is inferred for the data above 1 K. Assuming the presence of two superconducting BCS-like temperature dependent gaps we are able to describe the experimental data in the whole temperature range studied. The gap widths 2$\Delta_1/k_BT_c$ = 4.0 and 2$\Delta_2/k_BT_c$ = 2.5, and relative mole weights of the mole electronic heat coefficients $\gamma_1/\gamma_n$ = 0.7 and $\gamma_2/\gamma_n$ = 0.3 were fitted. From the analysis of the specific heat data we obtained characteristic thermodynamic parameters, like the specific heat jump at $T_c$, $\Delta C(T_c)/\gamma_n T_c$, electron-phonon coupling constant, $\lambda_{e-ph}$, upper $H_{c2}$ and thermodynamic critical $H_{c0}$ fields, Ginzburg-Landau penetration depth $\lambda_{L}$, coherence length $\xi$ and Ginzburg-Landau parameter $\kappa$. The obtained value of $\lambda_{e-ph}$, $N(E_F)$, $\Delta C(T_c)/\gamma_n T_c$ and 2$\Delta_0/k_BT_c$ can be accounted for intermediate electron-phonon coupling. We found a linear field dependence of the Sommerfeld coefficient at 0.4 K for fields below $\sim$ 2 T. The data of electronic band structure calculations show no overlapping between Mo-4d and Sb-sp orbitals and therefore, may support the proposed two-gap feature. In order to confirm the two-gap phenomenon in \MoSb, further studies using sophisticated techniques e.g. by means of spectroscopy techniques such as tunneling-, point-contact- or photoemission- spectroscopy and muon spin relaxation are planned.

\section{Acknowledgements}
The work at ILT\&~SR is supported by the grant No. N202 082 31/0449 of the
Ministry of Science and Higher Education in Poland.

\newpage
\begin{figure}
    \caption{Powder X-ray diffraction pattern of \MoSb. The observed (open circles) and calculated (solid line) profiles are shown on the top. The vertical marks in the middle are calculated positions of Bragg peaks. The line in bottom of the plot is the difference between calculated and observed intensities. The SEM image of the typical samples is shown in the inset.
    }
\label{Xray}
\end{figure}

\begin{figure}
\caption{ (Color Online) Total (dotted line) and partial densities (Sb1: dash-dotted, Sb2: dashed and Mo: solid line) of states of \MoSb.}
\label{DOS}
\end{figure}

 \begin{figure}
\caption{(Color Online)  The specific heat of \MoSb~divided by temperature as a function of square temperature. The dotted line is a guide to the eye.  The dashed line is a fit to Eq. \ref{eq.1}.  }
\label{CpT}
\end{figure}
\begin{figure}
\caption{(Color Online) a)  the electronic  specific heat in a logarithmic scale as a function of 1/\emph{T}. The dotted and dashed lines are theoretical ones (see text) The error bars are given with values of $\pm$ 20\% at 0.4 K and  $\pm$ 2\%  at 1.7 K.
b) Temperature dependence of the electronic entropy at 0 and 2 T.}
\label{Cel}
\end{figure}
\begin{figure}
\caption{(Color Online) The thermodynamic critical field $H_{c0}(T)$ as a function of temperature.  The dotted is a guide to the eye. The inset shows temperature dependence of the internal energy difference $\Delta U$ and entropy difference multiplied by the temperature $T\Delta S$. The dotted line is a guide to the eye.}
\label{Hc0}
\end{figure}

\begin{figure}
\caption{(Color Online) The temperature dependence of the specific heat divided by temperature  $C_p/T$ at several magnetic fields. The dotted line is a guide to the eye.}
\label{Cpfield}
\end{figure}

\begin{figure}
\caption{(Color Online) a) Upper critical field $H_{c2}$ vs. temperature. The dashed line represents the fit of the experimental data to $\mu_0H_{c2}(T)$ = $\mu_0H_{c2}(0)[1 - (T/T_c)^{1.29}]$. b) Field dependence of the Sommerfeld coefficient at 0.4 K. The dashed line illustrates a straight linearity of the $\gamma_{0.4 K} (H)$ dependence. }
\label{Hc2}
\end{figure}

\end{document}